\documentclass[pre,amsmath,amssymb]{revtex4}
\usepackage{bm}
\usepackage{graphicx}
\usepackage{amsthm}

\newcommand{\pacc}{p_\text{acc}}
\newcommand{\ptrial}{p_\text{tr}}

\usepackage{chemarr}

\begin{document}

\title{Absolute Monte Carlo estimation of integrals and partition functions}

\author{Artur B. Adib}
\email{adiba@mail.nih.gov}
\affiliation{
Laboratory of Chemical Physics, NIDDK, National Institutes of Health, Bethesda, Maryland 20892-0520, USA
}

\date{\today}

\begin{abstract}
Owing to their favorable scaling with dimensionality, Monte Carlo (MC) methods have become the tool of choice for numerical 
integration across the quantitative sciences.
Almost invariably, efficient MC integration schemes are strictly designed to compute ratios of integrals, their efficiency being intimately
tied to the degree of overlap between the given integrands.
Consequently, substantial user insight is required prior to the use of such methods, either to mitigate the oft-encountered lack of 
overlap in ratio computations, or to find closely related integrands of known quadrature in absolute integral estimation.
Here a simple physical idea---measuring the volume of a container by filling it up with an ideal gas---is exploited to design a new class 
of MC integration schemes that can yield efficient, absolute integral estimates for a broad class of integrands with simple transition matrices as input. 
The methods are particularly useful in cases where existing (importance sampling) strategies are most demanding,
namely when the integrands are concentrated in relatively small and unknown regions of configuration space 
(e.g. physical systems in ordered/low-temperature phases). Examples ranging from a volume with infinite support to the partition function of the 2D Ising model are 
provided to illustrate the application and scope of the methods.
\end{abstract}

\maketitle

\section{Introduction}

\begin{figure}
\begin{center}
\includegraphics[width=200pt]{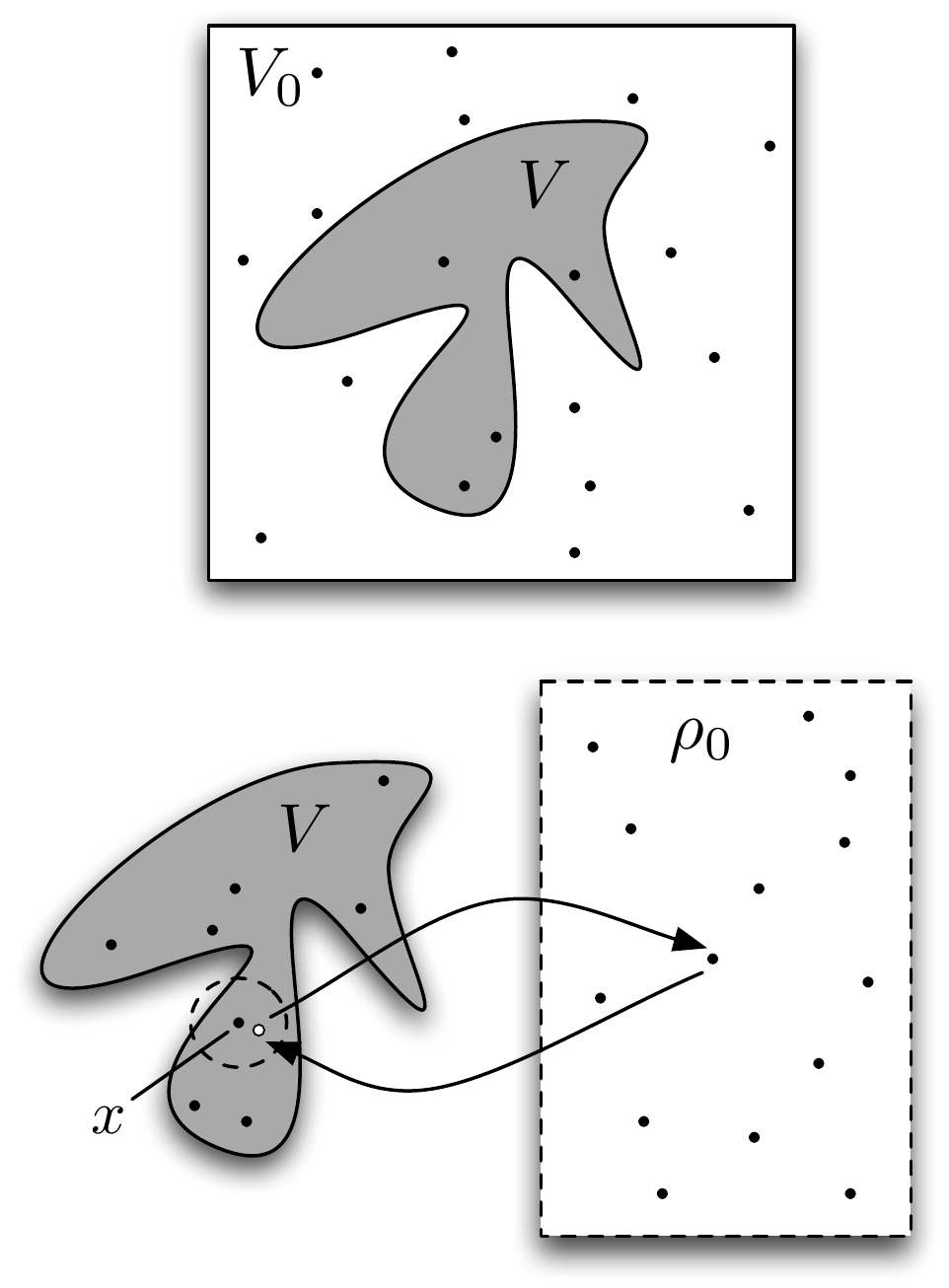}
\end{center}
\caption{ \label{fig:illustration}
Monte Carlo estimation of a volume $V$ (shaded region) by means of sampling from a reference volume $V_0$ (top), and
equilibration with a hypothetical, infinite reservoir of ideal gas particles at density $\rho_0$ (bottom).
In the former, one draws random points uniformly from $V_0$ and counts the fraction $f$ 
that lands in $V$, to obtain $V = V_0 f$. In the latter, one performs a grand-canonical Monte Carlo simulation \cite{frenkel02} at 
reservoir density $\rho_0$, and monitors the average number $\langle N \rangle$ of ideal gas particles in $V$; upon equilibration,
the density of particles in $V$ equals that of the reservoir, and thus $V = \langle N \rangle / \rho_0$. (In practice, due to the 
constraint $N\geq 1$, this formula for $V$ needs to be modified slightly; see Eq.~(\ref{eq:Z_1st})). 
Each particle corresponds to a point (``replica'') residing in $V$, and
attempted replica insertions/removals take place in the neighborhood (dashed circle) of an existing replica $x$, defined by the 
transition matrix $T(x'|x)$ of the method. A version of the algorithm with fixed number of replicas---possibly at different
temperatures---is also described in the text.
}
\end{figure}

To introduce and place the ideas of the present contribution in context, consider the paradigmatic problem of estimating the volume $V$ 
of a given region as shown in Figure~\ref{fig:illustration}. 
In its most rudimentary form, MC estimates of $V$ proceed using the ``hit-or-miss'' idea,
whereby the user designs a reference region of known volume $V_0$ that fully overlaps with $V$, and draws random points uniformly 
distributed in $V_0$ (cf. Fig.~\ref{fig:illustration}, top panel). An estimate of $V$ can then be obtained from $V/V_0 = f$, where $f$ is the 
fraction of such points that fall inside $V$. As is immediately apparent, the efficiency of this simple method hinges upon the amount of overlap 
between the volumes $V_0$ and $V$: the tighter $V_0$ bounds $V$, the more the hits and hence the better the quality of the 
estimate. Conversely, bounding volumes that have little overlap with $V$ give rise to more misses than hits and hence to large errors in the estimate.
This poses a major obstacle to the implementation
of such methods, especially in cases where it is difficult to guess the precise location of the volume and its boundaries, often leading to the 
design of an unnecessarily conservative (large) reference region, and hence to very inefficient estimates of $V$.

Despite its simplicity, the above example captures the central issues pertinent to integral estimation problems in general, for which 
more sophisticated methods exist \cite{liu-book,frenkel02,gregory-book}. Notably, in most efficient Monte Carlo techniques for normalizing constant 
or free energy estimation, a suitable family of intermediate integrands is required to 
interpolate between the two desired integrands (or between the available reference system and the integrand of interest, in
absolute integral estimation), the extent of overlap between the integrands dictating the efficiency of the methods much like in the 
example above \cite{bennett76,gelman98,jarzynski02-lncse}.
Such requirements become particularly daunting when the integrands in question are highly concentrated in unknown and disparate 
regions of configuration space, as is typically the case in most interesting physical problems, thereby demanding 
substantial user insight prior to the applicability of such methods.

In the present contribution, a new family of MC integration strategies that can greatly alleviate the demand for such
types of insights will be introduced.
As expressed by the central results underpinning these methods, Eqs.~(\ref{eq:Z_1st}) and (\ref{eq:Z_2nd}), the integrals of interest ($Z$, Eq.~(\ref{eq:Z})) are computed 
individually, thus fundamentally departing from the aforementioned alternatives that rely on ratios of similar integrals.
At the core of these new strategies is the use of integration spaces of enlarged dimensionality (``replicas''), 
a concept already widely invoked to speed up Markov chain simulations (parallel tempering \cite{deem05}, evolutionary Monte Carlo \cite{liu-book}, etc.), 
combined with the replacement of reference integrals by normalized transition matrices \cite{frenkel02} 
(also known as transition functions \cite{liu-book} or kernels \cite{gelman98} in the Markov chain literature).
Two main variants will be presented: the first, more physically intuitive, requires a fluctuating number of replicas
(Fig.~\ref{fig:illustration}, bottom panel); the second, more abstract but also more easily adaptable to existing replica simulations, 
uses a fixed number of replicas along with ``virtual'' insertions/deletions of replicas. These two variants are complementary to each 
other much like grand-canonical Monte Carlo (GCMC) and Widom's test particle insertion method in simulations
involving chemical potentials \cite{frenkel02}.

\section{Theory}

To see how in principle the simultaneous use of multiple configurations (replicas)
allows for the absolute computation of integrals, let us go back to the volume estimation
problem of Fig.~\ref{fig:illustration}. As illustrated in the bottom panel of that figure, the volume $V$ of interest can be 
estimated by equilibrating it with an infinite reservoir of ideal gas particles (replicas) of density $\rho_0$, 
and measuring the average number $\langle N \rangle$ of particles inside $V$, to obtain $V = \langle N \rangle/\rho_0$. 
This physical idea can be implemented computationally by means of a generalized form of the usual
grand-canonical Monte Carlo method \cite{frenkel02},
where the particle reservoir is at density $\rho_0$ (or, equivalently, at the corresponding chemical potential), 
and attempted particle insertions/deletions take place in the neighborhood of an existing particle rather than inside
a fixed volume that bounds $V$, that neighborhood being defined by a transition matrix $T(x'|x)$. These are the central ideas that motivated
the development of the two versions described below.

In general, we would like to estimate integrals of the type
\begin{equation} \label{eq:Z}
  Z = \int_\Omega dx \, \pi(x),
\end{equation}
where $\Omega$ is the support of the integral, and $\pi(x)$ is a positive-definite function (see below, however) of the $d$-dimensional vector $x$;
e.g. $\pi(x)=e^{-E(x)}$ for most physical problems with energy function $E(x)$.
Partition functions of discrete systems, i.e. $Z = \sum_{x \in \Omega} \pi(x)$, can be dealt with in an analogous fashion.
We are given a (normalized) transition matrix $T(x'|x)$, such as those routinely used in the trial part of the Metropolis 
algorithm \cite{frenkel02}; for example, $T(x'|x)$ could be a 
uniform probability distribution up to some distance $|x'-x|$ from $x$ (see dashed circle in Fig.~\ref{fig:illustration}),
or a Gaussian function centered about $x$. As with Metropolis and other Markov chain simulations, the width of these distributions 
can be chosen during a preliminary run of the simulation (see below).

For non-positive definite integrands, one can invoke the identity
\begin{equation} \label{eq:trick}
  Z = Z_{|\pi|} \, \langle \text{sgn}(\pi(x)) \rangle_{|\pi|}, 
\end{equation}
where $Z$ is defined by Eq.~(\ref{eq:Z}), $Z_{|\pi|} = \int_\Omega dx \, |\pi(x)|$, and the average of the sign function
of $\pi(x)$ is with respect to points $x$ sampled from $|\pi(x)|$. Provided they exist, 
both quantities on the right hand side of this identity are immediately available from the methods below.

\subsection{Varying number of replicas}

For the first version of the method, we would like to simulate a system 
of replicas in contact with a reservoir of ideal (non-interacting) replicas of specified density $\rho_0$, such that each 
replica $x_i$ in $\Omega$ independently samples the distribution $\pi(x_i)$. The corresponding grand-canonical partition function 
is thus
\begin{align}
  Q & = \sum_{N=1}^{\infty} \frac{\rho_0^N}{N!}\, \int_{\Omega} dx_1 \cdots \int_{\Omega} dx_N \, \pi(x_1)\cdots \pi(x_N) \nonumber \\
    & = \sum_{N=1}^{\infty} \frac{(\rho_0 Z)^N}{N!} = e^{\rho_0 Z} - 1. \label{eq:Q}
\end{align}
Note that, unlike the traditional case where $0 \leq N \leq \infty$, here the sum starts at $N=1$, as in the present method (see below)
replica insertions/deletions take place in the neighborhood of at least one existing replica. 
Provided we can simulate according to this partition function, the desired integral $Z$ can be found from the equality
\begin{equation} \label{eq:Z_1st}
  \frac{\langle N^2 \rangle}{\langle N \rangle} = 1 + \rho_0 Z,
\end{equation}
which follows straightforwardly from Eq.~(\ref{eq:Q}) by computing each moment of $N$ separately. (Alternatively, one can use
$\langle N \rangle = \rho_0 Z/(1-e^{-\rho_0 Z})$ or any other relationship between moments of $N$ and $Z$, and numerically solve the transcendental equation 
for $Z$; the question of which $Z$ estimator is more efficient will be left for future studies).

\begin{figure}
\begin{center}
\includegraphics[width=250pt]{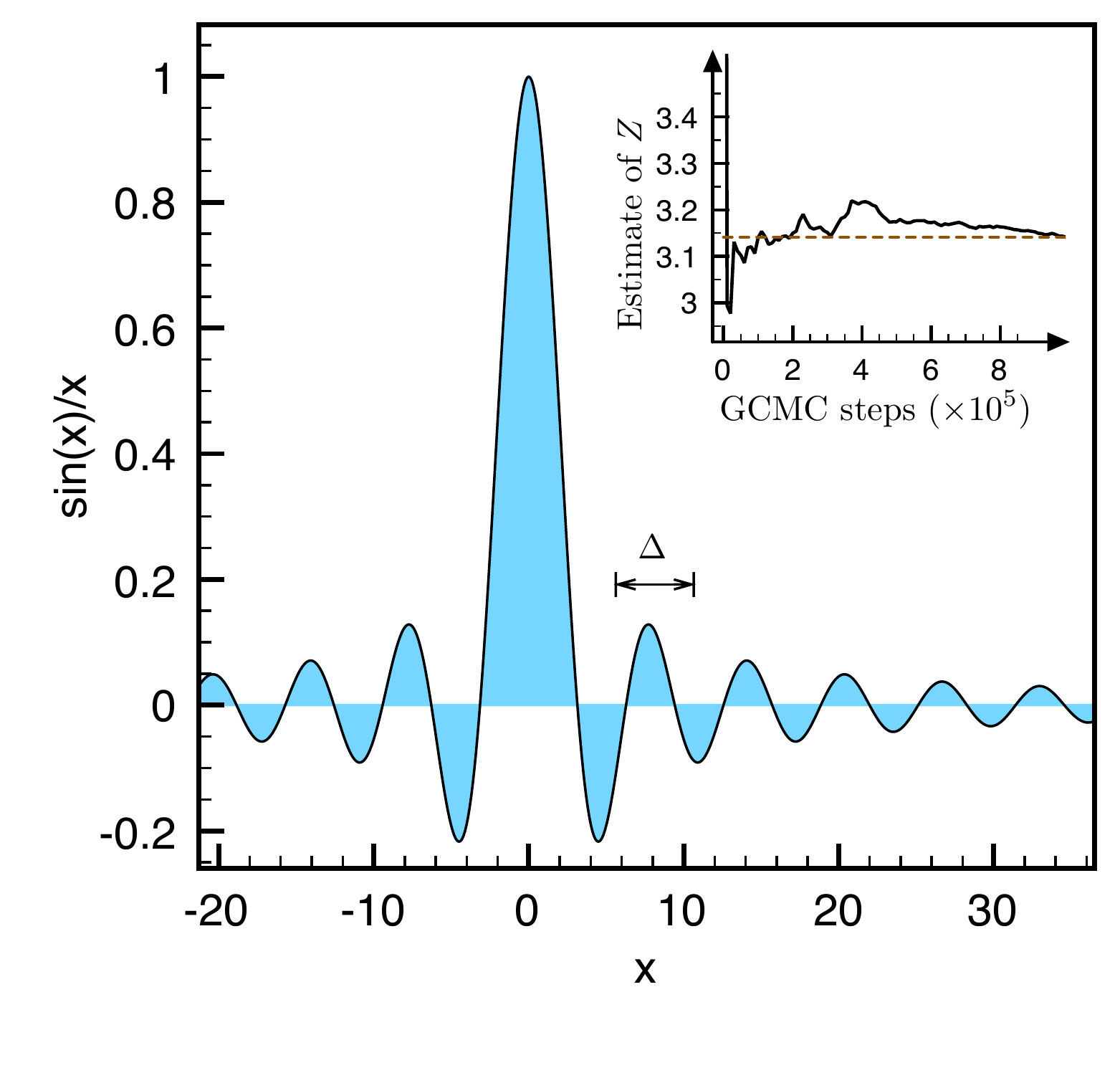}
\end{center}
\caption{ \label{fig:sinx}
An illustrative non-positive definite integrand, $\pi(x) = \sin(x)/x$.
The scale in the center of the graph corresponds to the parameter $\Delta = 5$ in the uniform transition matrix $T(x'|x) = 1/\Delta$ for $|x'-x| < \Delta/2$  
(zero otherwise) adopted for the results shown in the inset.
{\em Inset:} Running estimate of $Z = \int_{-\infty}^{\infty} dx \, \pi(x)$ using Eq.~(\ref{eq:trick}), 
where $Z_{|\pi|}$ is estimated via the replica gas method with varying number of replicas, Eq.~(\ref{eq:Z_1st}).
The mean sign function of $\pi$ required by this last equality, $\langle \text{sgn}(\pi(x)) \rangle_{|\pi|}$, is also obtained from this run, 
by averaging $\text{sgn}(\pi(x))$ over all replicas $x$ during the simulation.
The dashed red line is the exact result $Z = \pi = 3.141592...$. For this example, $\rho_0 = 1$, and $10$ ordinary Monte Carlo moves per replica are performed 
between every attempted insertion/deletion (GCMC step). The ensuing number of replicas fluctuated about $N = 10$.
}
\end{figure}

An algorithm that samples according to the above grand-canonical partition function goes as follows.
At the beginning of the algorithm we are given at least one point $x_1$ that belongs to $\Omega$; let us
assume the general case where we already have $N$ replicas in $\Omega$, and let us denote the vector of coordinates of the replicas
by $x^N \equiv ( x_1, \ldots, x_N )$. We then decide whether to insert or remove a replica, 
typically---but not necessarily (see Methods section)---with equal probability. If the decision was an insertion, 
we sample a new replica coordinate $x_{N+1}$ from $T(x_{N+1}|x_i)$, where $x_i$ is a randomly chosen coordinate from the existing $N$,
and accept its insertion with probability
\begin{equation} \label{pacc1}
  \pacc( x^{N+1} | x^N ) = \min\left\{ 1, \frac{N}{N+1} \, \frac{\rho_0 \, \pi( x_{N+1} )}{\sum_{i=1}^N T(x_{N+1}|x_i)} \right\},
\end{equation}
except when $x_{N+1}$ lies outside $\Omega$, in which case an immediate rejection takes place.
Similarly, if the decision was to try a deletion, we randomly pick a replica $x_j$ from the existing $N$, and
accept its deletion with probability
\begin{equation} \label{pacc2}
  \pacc( x^{N-1} | x^N ) = \min\left\{ 1, \frac{N}{N-1} \, \frac{\sum_{i \neq j}^N T(x_j|x_i)}{ \rho_0 \, \pi(x_j)} \right\},
\end{equation}
where $x^{N-1}$ is $x^N$ excluding $x_j$, and the sum over $i$ excludes $i=j$. 
An exception is the case where only one replica remains, which is always rejected.
A proof that this algorithm samples according to Eq.~(\ref{eq:Q}) follows by detailed balance (see Methods section). 
When the replicas correspond to particles inserted uniformly in a fixed volume $V$, i.e. $T(x'|x) = 1/V$,
the above acceptance probabilities reduce to those of Ref.~\cite{frenkel02} (with the due mappings between $\rho_0$ and chemical potential,
and between $\pi(x)$ and the Boltzmann factor). Of course, it is also possible to perform ordinary $\pi$-preserving
Monte Carlo moves on each replica before attempted insertion/deletions \cite{frenkel02} (Fig.~\ref{fig:sinx} uses this idea).

By repeating the above procedure a number of times, the simulation will eventually equilibrate, and the number of
replicas will fluctuate about its mean value $\langle N \rangle$. If the equilibration is too slow, i.e. too few
replica insertions/deletions are accepted, the width of the distribution $T(x'|x)$ about $x$ can be adjusted accordingly
during a preliminary run, in a fashion analogous to what is done in Metropolis Monte Carlo to keep the rate of 
accepted trial moves within a reasonable range \cite{frenkel02}. Likewise, if the number of replicas starts to grow
beyond one's computational capabilities, or diminish until it hardly departs from unity, $\rho_0$ can be adjusted
so that $\langle N \rangle$ is a reasonable number consistent with one's computing power. In practice, for many-particle systems
with extensive free energies (i.e. $Z \sim e^{N}$), it is best to write $\rho_0 = e^{\mu_0 N}$ and adjust $\mu_0 < 0$ instead.

Note that the present algorithm generalizes standard grand-canonical simulation \cite{frenkel02}
in two crucial ways. First, it inserts and removes entire replicas of the system of interest as opposed to individual particles of a many-body system. This 
conceptual difference is essentially what allows one to relate moments of $N$ to the partition function of the system of interest (Eq.~(\ref{eq:Z_1st})).
Second, the replicas are introduced in the neighborhood of an existing replica instead of inside a fixed region, as prescribed by the transition matrix. 
This allows for efficient simulation
when the integrand of interest is sharply peaked about unknown regions of configuration space (cf. Fig.~\ref{fig:illustration2}). 
Although the use of arbitrary transition matrices in grand-canonical simulations is known in the
mathematical literature \cite{moller-book}, to our knowledge the use of such ideas for integral/partition function estimation
is new.

\subsection{Fixed number of replicas}

The second version of the replica gas method introduces two important advantages.
First, the number of replicas is constant as opposed to fluctuating, making it more convenient for parallel computing architectures, 
and second the replicas can be simulated at different temperatures. 
These features also make the method easily implemented in existing parallel tempering (replica exchange) simulations,
thereby benefiting from the greatly enhanced equilibration rates of these simulations \cite{deem05}.
The integrals of interest can then be estimated by computing two separate averages involving 
both the integrand $\pi(x)$ and the transition matrix $T(x'|x)$, as described below.

To introduce the method in its simplest form, let us first assume that only two replicas exist ($N=2$), 
each of them independently sampling the distributions $\pi$ and $\tilde{\pi}$, via e.g. Metropolis.
The integral of interest is $Z$, as in Eq.~(\ref{eq:Z}), and $\tilde{Z} = \int_\Omega dx \, \tilde{\pi}(x)$ is an auxiliary 
integral; the auxiliary distribution $\tilde{\pi}$ is arbitrary (for example, it could be $\pi$ itself), but in 
typical applications it corresponds to $\pi$ at a higher temperature, i.e. $\tilde{\pi}(x) = \pi^\beta(x)$, where $0 < \beta < 1$.
Then the following identity holds:
\begin{align}
  Z & = \frac{\int_\Omega dx [\tilde{\pi}(x)/\tilde{Z}] \int_\Omega dx' \, T(x'|x) \cdot \pi(x')}{ \int_\Omega dx [\tilde{\pi}(x)/\tilde{Z}] \int_\Omega dx' [\pi(x')/Z] \cdot T(x'|x)} \nonumber \\
    & \equiv \frac{\left\langle \pi(x') \right\rangle_{\tilde{\pi},T} } { \left\langle T(x'|x)  \right\rangle_{\tilde{\pi},\pi} }, \label{eq:Z_2nd}
\end{align}
where the average $\langle \mathcal{O}(x,x') \rangle_{f,g}$ of an observable $\mathcal{O}(x,x')$ means that configurations $x$ ($x'$) are sampled
from the distribution $f$ ($g$). Thus, the numerator in the above result requires $x$ to be sampled from 
$\tilde{\pi}(x)$ while $x'$ is sampled from $T(x'|x)$ (``virtual replica insertion,'' in analogy with Widom's method \cite{frenkel02}), and for each such pair of configurations
one computes the value of $\pi(x')$. Likewise, for the average in the
denominator, $x$ is sampled from $\tilde{\pi}(x)$ while $x'$ is sampled from $\pi(x')$, and for each pair one evaluates $T(x'|x)$ (``virtual replica deletion'').
In the limit of infinite samples, the ratio of these two averages converges to $Z$ as expressed by Eq.~(\ref{eq:Z_2nd}).

For simulations with multiple replicas at different temperatures ($\beta_1,\ldots,\beta_N$), one can simply combine (i.e. sum) the above result for each pair of replicas.
Thus, for the Ising model results in Fig.~\ref{fig:ising2d}, the equation
\begin{equation} \label{eq:Z_2nd_manybeta}
  Z(\beta_i) = \frac{\sum_{j\neq i} \left\langle e^{-\beta_i E(x')} \right\rangle_{\beta_j,T} } { \sum_{j\neq i} \left\langle T(x'|x)  \right\rangle_{\beta_j,\beta_i} }
\end{equation}
was used. The sums run over each replica $j$ at temperature $\beta_j$, except $j=i$. The energy function $E(x)$ for a spin configuration $x$
is the usual Ising model function $E(x) = -\sum_{\langle k,l \rangle} x_k x_l$ with periodic boundary conditions \cite{krauth-book}.
The transition matrix adopted $T(x'|x)$ generates a new spin configuration $x'$ by flipping each spin of $x$ with probability $p_\text{flip}$.
Thus, $T(x'|x) = (p_\text{flip})^{\parallel x' - x \parallel} (1-p_\text{flip})^{N-\parallel x'- x\parallel}$, 
where $\parallel x' - x \parallel$ is the Hamming distance (number of spin mismatches) between the configurations $x$ and $x'$, 
and $N$ is the total number of spins.

\section{Results and Discussion}

\begin{figure}
\begin{center}
\includegraphics[width=250pt]{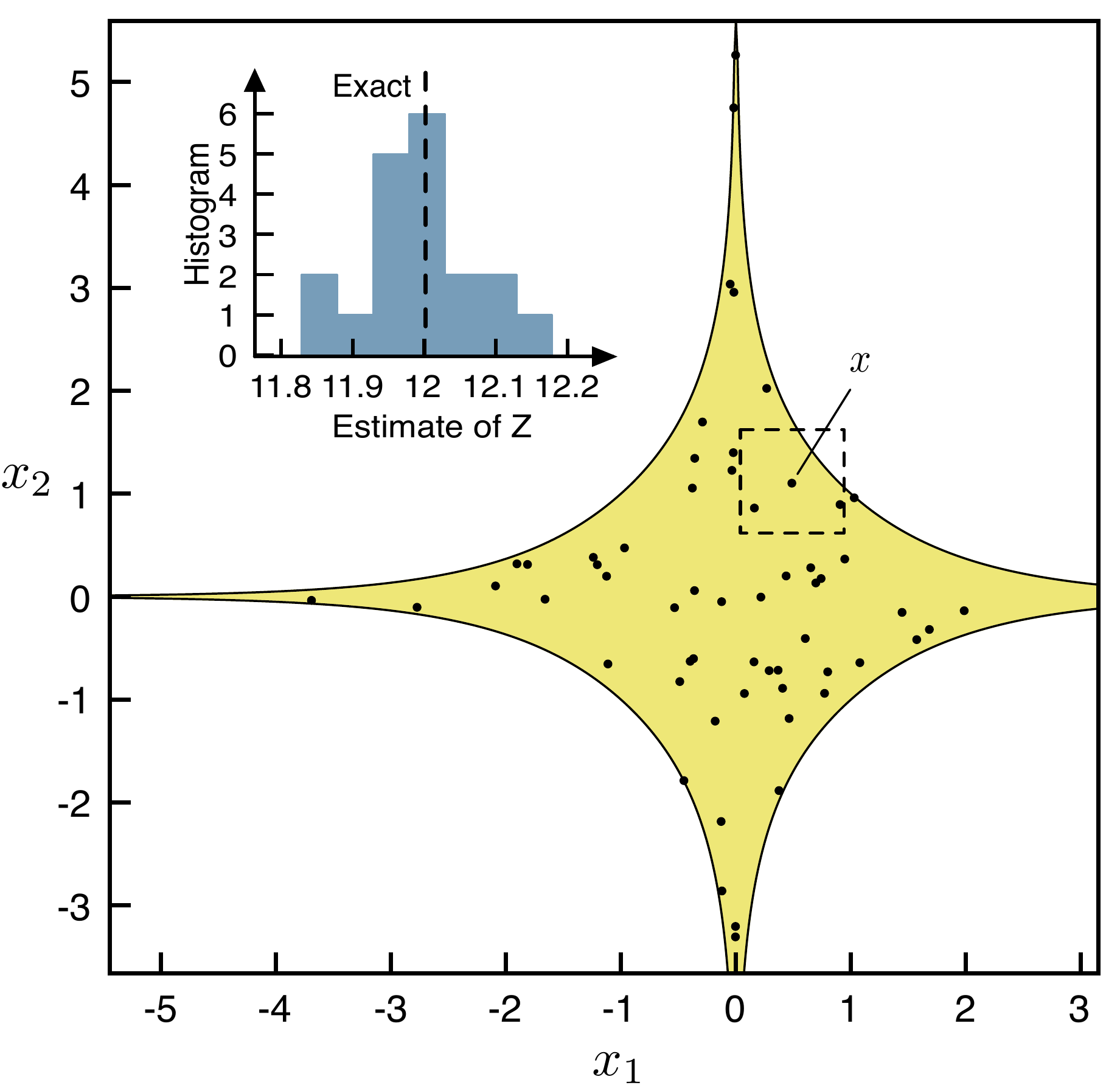}
\end{center}
\caption{ \label{fig:vol2d}
Estimation of a two-dimensional volume $Z$ (yellow region) using the replica gas method with varying number of replicas, Eq.~(\ref{eq:Z_1st}).
The volume $Z$ is defined by the region $|x_2| \leq e^{-|x_1|+1}$ for $|x_1|>1$, and $|x_2| \leq 1-\ln|x_1|$ for $|x_1|\leq 1$. The points correspond to 
the replica configurations at the end of one simulation, 
and the dashed square defines the boundaries of the adopted transition matrix $T(x'|x)$ (uniform distribution, each side of length unity) 
for the particular configuration $x$ shown. 
{\em Inset:} Histogram of $20$ independent estimates of $Z$ using the replica gas method with $10^6$ attempted insertions/deletions, and $\rho_0 = 5$.
The exact value of $Z$, obtained by analytic quadrature, is $Z=12$.
}
\end{figure}

For illustrative purposes, the results of a volume estimation problem in two spatial dimensions using the version with varying number of replicas are 
reported in Figure~\ref{fig:vol2d}.  This example was chosen due to its infinite support,
a property that would render the use of importance sampling methods difficult, as they would require 
the design of a non-trivial reference volume $V_0$ with similar support and known quadrature (recall that in principle we do not know where 
the integrand is concentrated, or where its boundaries are). 
The present method performs well in such circumstances without any prior information concerning the support 
of the integrand, by using a simple uniform transition matrix $T(x'|x)$ (Fig.~\ref{fig:vol2d}, dashed square).

As an application to integrals more general than simple volumes, in Fig.~\ref{fig:sinx} a representative estimate of
$Z = \int_{-\infty}^{\infty} dx \sin(x)/x$ is shown. Note that this integrand is non-positive definite, so Eq.~(\ref{eq:trick}) was used.
The quantities on the right hand side of that equation were estimated using the varying number of replicas version of the method,
with the positive-definite integrand $|\pi(x)| = |\sin(x)/x|$.

\begin{figure}
\begin{center}
\includegraphics[width=250pt]{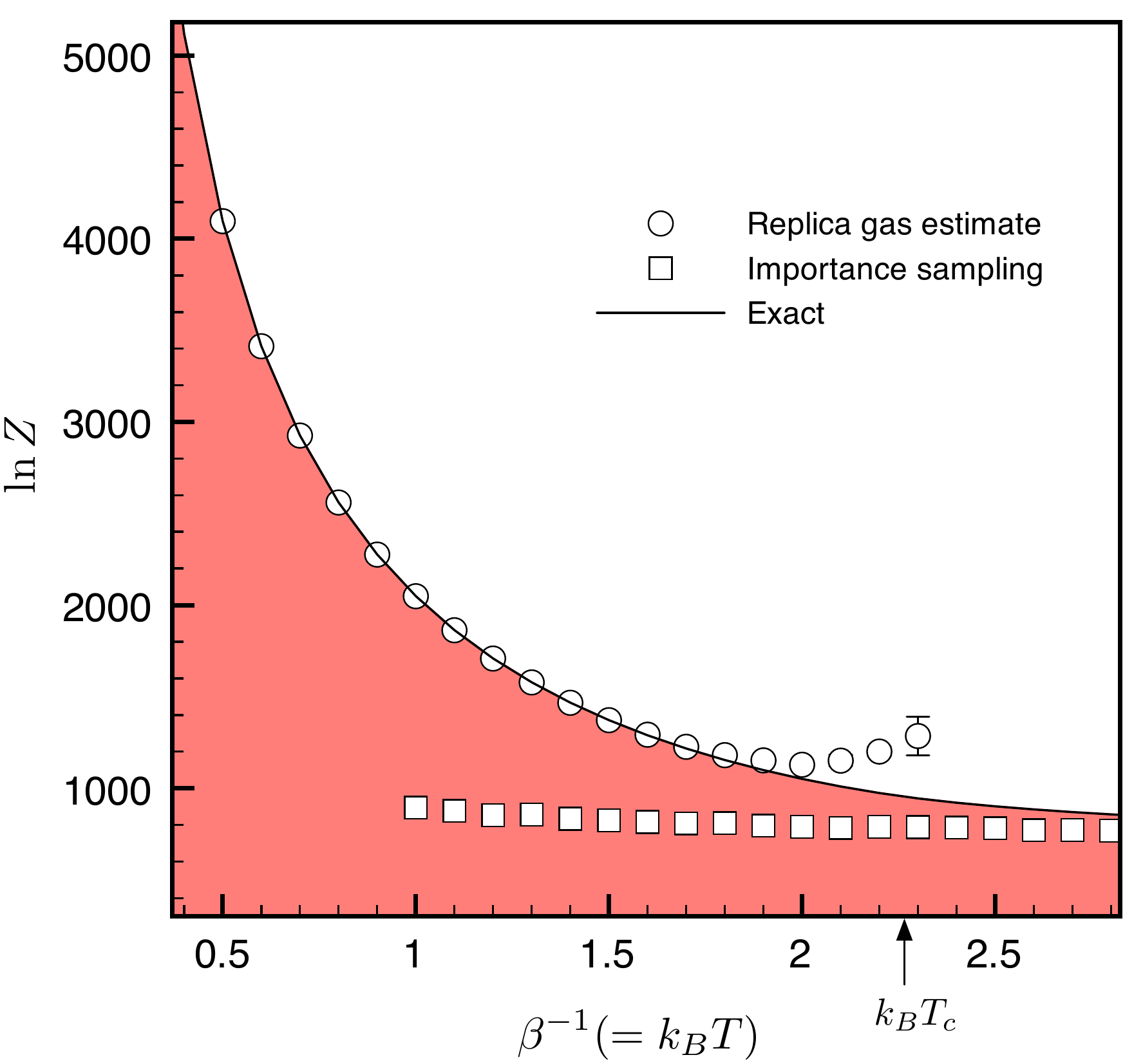}
\end{center}
\caption{ \label{fig:ising2d}
Natural logarithm of the partition function $Z$ of the two-dimensional Ising model with $N = 32 \times 32$ spins, 
according to the replica gas method (circles), and the exact Kaufman formula \cite{krauth-book} (dashed curve).
The replica gas results were obtained using the fixed number of replicas version, Eq.~(\ref{eq:Z_2nd_manybeta}), with $20$ 
replicas at the temperatures corresponding to the data points shown (similar results are obtained with varying number of replicas, see text), 
with error bars indicating the standard deviation of 8 independent runs. Importance sampling results were obtained using 
the ideal (non-interacting) spin partition function $Z_\text{id} = 2^N$ and
$Z_\text{Ising} / Z_\text{id} = \langle e^{-\beta E_\text{Ising}(x)} \rangle_\text{id}$,
with $10^5$ independent configurations $x$ sampled from the ideal reference system (increasing this number to $10^6$ does not lead to appreciable
changes in the results). For disordered states (i.e. $k_B T$ close to or higher than the
critical temperature $k_B T_c = 2.269$), the partition function is no longer dominated by a small fraction of the
configuration space, and the replica gas method converges more slowly with the adopted (local) transition matrix (see also Fig.~\ref{fig:illustration2}).
The replica exchange simulation took $10^5$ MC steps, with an attempted exchange every $100$ steps, where each 
MC step is a simple spin flip. For the transition matrix sampling, $p_\text{flip} = 1/N$ (cf. discussion after Eq.~(\ref{eq:Z_2nd_manybeta})).
}
\end{figure}

\begin{figure}
\begin{center}
\includegraphics[width=250pt]{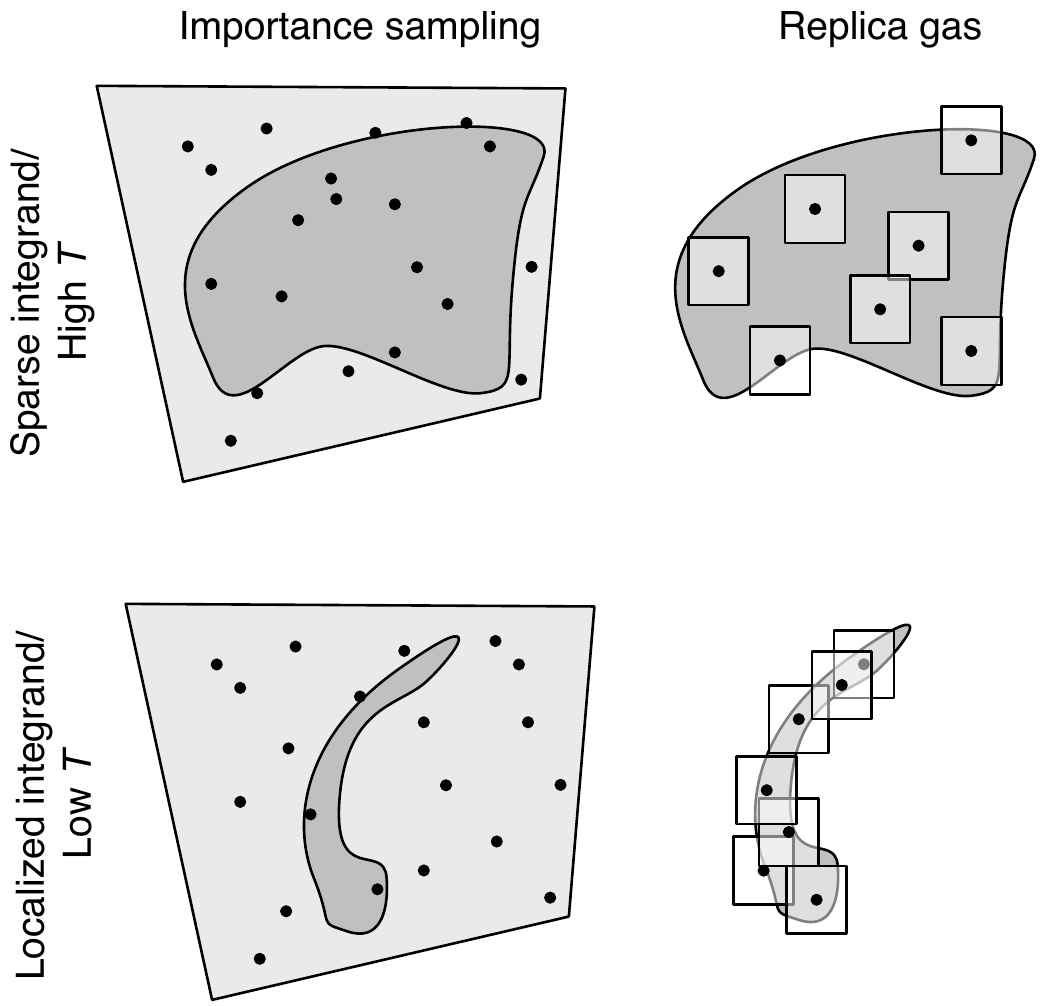}
\end{center}
\caption{ \label{fig:illustration2}
Comparison of the merits of importance sampling (left) and replica gas (right) methods for sparse (top) and localized (bottom) integrands.
Integrands are represented by their densest regions (shaded curvy shapes).
For physical systems, sparse integrands correspond to Boltzmann factors at high temperatures, while at low temperatures the integrands tend to be 
localized in a small fraction of configuration space (e.g. magnetized spin systems, crystals, proteins in their native state, etc). In importance sampling, one typically
has at their disposal a general-purpose sparse reference system (polygon on the left) such as an ideal gas, which is generally sufficient to ensure proper sampling 
at high temperatures, but not at lower ones.
Conversely, in replica gas methods one typically has at their disposal a local transition matrix (small boxes on the right) that is generally sufficient
to sparingly ``cover'' the integrand of interest (cf. efficiency criteria (a) and (b) discussed in the text) at low temperatures,
but not at higher temperatures.
}
\end{figure}

In Figure~\ref{fig:ising2d} the partition function of the two-dimensional Ising model is computed to demonstrate the version with fixed number of replicas
(similar results are obtained with the non-fixed version).
At low temperatures, i.e. ordered states, the replicas are densely localized about 
the spin-up and spin-down states, and hence a local transition matrix $T(x'|x)$ is sufficient to ensure efficient convergence of the averages.
Conversely, for higher temperatures close to the disordered state and above, the relevant configuration space---and hence the spread of 
the replicas---grows beyond the reach of the local transition matrix adopted, thus causing the averages to converge more slowly (cf. right side of 
Fig.~\ref{fig:illustration2}).

To understand such convergence issues in more detail, consider for simplicity Eq.~(\ref{eq:Z_2nd}) when $\tilde{\pi} = \pi$
(see below for the version with varying number of replicas).
In order for the averages in Eq.~(\ref{eq:Z_2nd}) to converge efficiently, the transition matrix $T(x'|x)$ has to be such that:
\begin{itemize}
\item[(a)] Most configurations $x'$ sampled from $T(x'|x)$ fall in typical regions of $\pi$ for any typical configuration $x$ sampled from $\pi$ 
(so that the numerator is not dominated by those rare configurations with high values of $\pi(x')$);

\item[(b)] Most configurations $x$,$x'$ sampled from $\pi$ fall in typical regions of $T(x'|x)$
(so that the denominator is not dominated by those rare events that cause $T(x'|x)$ to be of appreciable value).
\end{itemize} In the Ising model example of Fig.~\ref{fig:ising2d}, where $T(x'|x)$ typically flips only a few spins of $x$, the low temperature estimates 
converge faster as typical spin configurations only differ by a few spins, thereby satisfying both requirements, while at higher temperatures close
to $T_c$ and above, any two typical spin configurations differ by a substantial number of spins, and hence the requirement (b) is violated. Adding more replicas,
increasing the value of $p_\text{flip}$ for higher temperatures, or using non-local transition matrices (such as those of cluster algorithms \cite{krauth-book}) 
can alleviate the problem, but such issues will be left for future studies.

Note that analogous convergence issues arise in the version with varying number of replicas. Indeed, as can be seen by inspection of Eqs.~(\ref{pacc1}) and (\ref{pacc2}), 
the acceptance probabilities for insertion and deletion are affected by the choice of $T(x'|x)$ much like the averages in Eq.~(\ref{eq:Z_2nd}) are affected by
criteria (a) and (b) above. (A separate issue is how well the replicas explore the energy landscape. In the fixed number of replicas version, 
different temperatures are used to overcome energy barriers. Although replica exchange operations can be combined with the varying number of replicas method, 
as a proof of concept for the convergence issues above, it suffices to start a population of replicas that populate spin-up and spin-down states equally).

As illustrated by the above example, the most attractive use of the present methods lies in problems where
the integrand is sufficiently localized, so that a general-purpose, local transition matrix can be used to yield efficient
results with moderate numbers of replicas. These are precisely the problems for which existing importance sampling-based methods are most demanding,
and thus these methods can be seen as complementary to each other (see Fig.~\ref{fig:illustration2}). 
It should be noted that the so-called ``flat histogram'' Monte Carlo 
methods \cite{wang01a,shell07} are also able to bypass some of the difficulties with importance sampling strategies in some cases, 
especially for discrete systems. However, the required human input and scope of such integration methods are rather different: they require the existence 
and knowledge of suitable order parameters and their ranges (this being particularly difficult for entropic problems, such as that of Fig.~\ref{fig:vol2d}), 
knowledge of ground state degeneracies, and for continuum systems suffer from systematic errors due to discretization schemes, although
attempts to alleviate some of these problems have been put forward \cite{troster05}.

In summary, the present contribution has introduced two variants of a novel Monte Carlo strategy for estimating integrals and partition functions. 
Both versions can be seen as complementary to existing importance sampling or free energy methods \cite{liu-book,gelman98,jarzynski02-lncse}, in that 
their utility is generally best when the integrands are concentrated in relatively small and unknown regions of configuration space.
Both continuum and discrete systems are equally amenable to their use. 
By shifting focus from importance sampling functions to transition matrices, it is expected that these methods will
encourage a change of paradigm in Monte Carlo integration.

\section{Methods}

In this section it will be shown that Eqs.~(\ref{pacc1}) and (\ref{pacc2}) satisfy the detailed balance condition
\begin{equation}
  p( x^N ) \cdot \ptrial( x^{N+1} | x^N ) \cdot \pacc( x^{N+1} | x^N )
    = p( x^{N+1} ) \cdot \ptrial( x^N  | x^{N+1} ) \cdot \pacc( x^N | x^{N+1} ). \label{detbalance}
\end{equation}
According to the grand-canonical partition function Eq.~(\ref{eq:Q}), the probability of observing the 
microstate $x^N$ is given by
\begin{equation} \label{labeled}
  p(x^N) \propto \frac{\rho_0^N \pi(x_1) \cdots \pi(x_N)}{N!},
\end{equation}
where the proportionality constant does not depend on the replica coordinates or $N$. Note carefully the difference between the distribution
of the labeled microstate $x^N$, corresponding to replica with label ``1'' being at $x_1$, ``2'' at $x_2$, etc, and that of the unordered set of 
coordinates $\{ x^N \} = \{ x_1, \ldots, x_N \}$, corresponding any replica being at $x_1$, another arbitrary replica at $x_2$, etc. This
probability is given by
\begin{equation} \label{unlabeled}
  p(\{ x^N \}) = \sum_P p(x^N) \propto \rho_0^N \pi(x_1) \cdots \pi(x_N),
\end{equation}
where $\sum_P$ is the sum over all possible permutations of $x_1,\ldots,x_N$. Since the replicas
are indistinguishable, we have $p(\{ x^N \}) = N! \, p(x^N)$, as in above. Of course, it is possible to use either description (labeled 
or unlabeled), provided the correct probability distributions are used (Eq.~(\ref{labeled}) or Eq.~(\ref{unlabeled}), respectively).
In this section, following \cite{filinov69}, we will only show the proof using labeled states, i.e. Eq.~(\ref{labeled}). 
It is a simple exercise to modify the development below for the unlabeled case; the acceptance probabilities, of course, are unchanged.

Our acceptance probabilities are of the Metropolis-Hastings type, which by construction satisfy detailed balance. In the present
notation, the formulas are
\begin{equation} \label{methastings}
  \pacc( x^{N+1} | x^N ) = \text{min} \left\{ 1, \frac{\ptrial( x^N | x^{N+1} )}{\ptrial( x^{N+1} | x^N )} \frac{p( x^{N+1} )}{p( x^N )} \right\},
\end{equation}
and analogously for $\pacc( x^N | x^{N+1} )$. According to the insertion/deletion algorithm described before Eq.~(\ref{pacc2}), the trial probabilities 
for going between the states $x^N = (x_1, \ldots, x_N)$ and $x^{N+1} = ( x_1, \ldots, x_N, \xi )$ are given by
\begin{equation} \label{trial}
  ( x_1, \ldots, x_N ) \xrightleftharpoons[(1-q) \cdot \frac{1}{N+1}]{q \cdot \frac{1}{N+1} \cdot \sum_{i=1}^N \frac{1}{N} T(\xi|x_i) } ( x_1, \ldots, x_N, \xi ),
\end{equation}
where $\ptrial( x^{N+1} | x^N )$ is given by the expression above the arrows, and $\ptrial( x^N | x^{N+1} )$ by the one below them.
In the insertion trial probability,  $q = 1/2$ is the probability to try an insertion as opposed to a deletion, 
$1/(N+1)$ is the probability that the new coordinate $\xi$ is inserted
in a given slot of the vector $x^{N+1}$ (in the above case, the last slot), $1/N$ is the probability to pick coordinate $x_i$ as reference, 
and $T(\xi|x_i)$ is the probability to sample the candidate position $\xi$ given the chosen reference. In the deletion trial probability,
$(1-q) = 1/2$ is the probability to try a deletion, and $1/(N+1)$ is the probability that the replica at
$\xi$ will be chosen for attempted removal among the existing ones. 
Plugging these trial probabilities in Eq.~(\ref{methastings}), we obtain Eq.~(\ref{pacc1}) (an analogous
procedure gives Eq.~(\ref{pacc2})). The case where $q \neq 1/2$ can be easily taken care of by modifying the acceptance probabilities accordingly.

Alternatively, detailed balance can be directly proven by plugging the trial probabilities in Eq.~(\ref{trial}) and the acceptance probabilities given by 
Eqs.~(\ref{pacc1}) and (\ref{pacc2}) into Eq.~(\ref{detbalance}).

\begin{acknowledgments}
The author would like to thank Attila Szabo, Gerhard Hummer, and David Minh for discussions and suggestions. This research was supported by the Intramural Research Program of the NIH, NIDDK.
\end{acknowledgments}

%
%

%
%

%
%

%
%

\end{document}